\documentstyle[12pt,epsfig]{article}                 
\def\f {\beta}
\def\a {\beta_A}

\def\t {N_{\tau}}

\begin{document}
\begin{titlepage}
\vskip1.5cm

\begin{center}
{\Large {\bf $Z_2$ Monopoles, Vortices and the Universality of the 
SU(2) Deconfinement Transition }} 
\\[15mm]

{\large Rajiv V. Gavai$^{a,}$\footnote{E-mail:gavai@tifr.res.in}
and Manu Mathur$^{b,}$\footnote{E-mail:manu@boson.bose.res.in}}  \\[5mm]
{\em $^a$Theoretical Physics Department, Tata Institute of Fundamental
Research, \\
Homi Bhabha Road, Mumbai 400005, India} \\[2mm]
{\em $^b$S. N. Bose National Centre for Basic Sciences, JD Block, \\
Sector III, Salt Lake City, Calcutta 700091, India. } \\[20mm]
\end{center}

\begin{abstract}
We investigate the effect of $Z_2$ magnetic monopoles and vortices on 
the finite temperature deconfinement phase transition in the 
fundamental - adjoint $SU(2)$ lattice gauge theory.  In the limit of complete 
suppression of the $Z_2$ monopoles, the mixed action for the $SU(2)$ theory 
in its Villain form is shown to be self-dual under the exchange of the 
fundamental and adjoint couplings. By further suppressing the $Z_2$ vortices 
we show that the extended model reduces to the Wilson action 
with a modified coupling. 
The universality of the $SU(2)$ deconfinement 
phase transition with the Ising model is therefore expected to remain 
intact in the entire plane of the fundamental-adjoint couplings in the 
continuum limit. The self-duality arguments related 
to the suppression of $Z_2$ monopoles are also 
applicable to the Villain form of mixed action for the $SU(N)$ theory
with $Z_{N}$ magnetic monopoles.  

\end{abstract} 

\end{titlepage}

\begin{center}
\bf 1. INTRODUCTION \\
\end{center}
\bigskip

Confinement of non-abelian color degrees of freedom is widely regarded as an 
outstanding problem in physics. Condensation \cite{thooft} of certain 
magnetic monopoles is a plausible mechanism to explain it. Quantizing the 
theory on a discrete space-time lattice, one hopes to shed more light on 
the inherently non-perturbative phenomenon of quark confinement. 
Although, this quantization procedure is not unique, different ways 
of defining the gauge theories on lattice are expected to lead to the 
same physics when the continuum limit of vanishing lattice spacing
is taken.  In the strong coupling region, however, where the coupling 
constant and lattice spacing are relatively large, it is expected that 
lattice artifacts will usually affect the physics.  Often such changes
are purely quantitative in nature: the string tension or the hadron
spectra exist in the strong coupling expansion but the dependence
of their dimensionless ratios on the coupling changes 
as one goes in to the asymptotic scaling region.
By incorporating higher order corrections in the coupling constant,
one can hope for a larger scaling region where the quantitative 
differences due to artifacts  disappear with decreasing lattice 
spacing. However, the lattice artifacts may affect physics even 
{\bf qualitatively}. The simplest known and 
well understood example is that of compact U(1) lattice gauge theory.  
It has magnetic monopoles which arise due to the compact
nature of the gauge fields. In the strong coupling region their
condensation gives rise to a confining phase with a linearly rising 
potential between electric charges: V(r) $\propto$ ~r. This is 
well known as the dual Meissner effect. The theory undergoes a first 
order phase 
transition at intermediate value of electric charge. Beyond this 
transition, the magnetic monopoles become irrelevant  and
the standard Coulomb potential, V(r) $\propto$ ~r$^{-2}$,
is recovered. Thus physical laws with a large lattice spacing can be 
drastically different in nature than in the continuum limit. 
To the best of our knowledge, no such example  exists
in non-abelian lattice gauge theories.
We will discuss below a similar phenomenon in SU(2) lattice
gauge theory which is due to $Z_2$ magnetic monopoles and 
$Z_2$ electric vortices. 

In non-abelian lattice gauge theories, many quantitative tests of the 
universality of the continuum limit have been carried out in the past. 
In particular, the fundamental-adjoint SU(2) action\cite{BhaCre} 
has been extensively studied \cite{HalSch,DasHel,AlbFly,OgiHor} in this 
context due to its rich phase structure, shown in Fig. 1 by solid 
lines of bulk phase transitions \cite{BhaCre,gavbulk}.
We argued in Ref. \cite{us1} that the phase diagram in Fig. 1 with only 
bulk boundaries was incomplete: it must have lines of deconfinement
phase transition for different $\t$. Subsequently, we located the 
the deconfinement transition lines\cite{us1,us2,us3} in Fig. 1 using 
the behavior of the corresponding order parameter.  
Along the Wilson axis ($\a =0.0$), the deconfinement transition has been 
very well studied. Monte Carlo simulations have established $i)$ a scaling 
behavior\cite{FinHelKar} for the critical coupling, and thus an approach 
to the continuum limit, and $ii)$ a clear second order deconfinement 
phase transition with critical exponents\cite{EngFin} in very good 
agreement with those of the three dimensional Ising model.  These provided 
an explicit verification of the Svetitsky and Yaffe\cite{SveYaf} 
universality conjecture.

\begin{figure}[htbp]\begin{center}
\epsfig{height=10cm,width=10cm,angle=-90,file=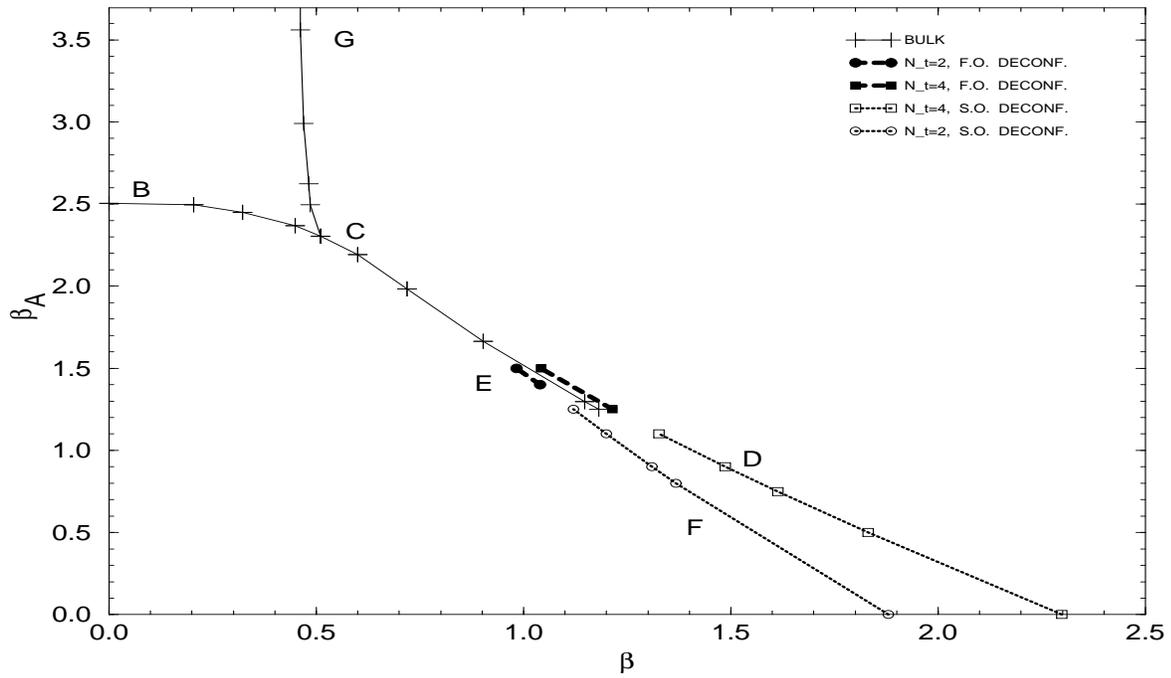}
\caption{The phase diagram of the mixed $SU(2)$ lattice gauge theory.  
The solid lines are from Ref. \cite{BhaCre,gavbulk}. The dotted(thick-dashed) 
lines with hollow(filled) symbols are the second(first)  order 
deconfinement phase transition lines\cite{us1,us2,us3} on $\t$ =2 
(circles) and 4 (squares) lattices. }
\end{center}\end{figure}

The continuum physics should not depend upon the group 
representation chosen to describe the dynamical variables. 
Therefore it is expected that the adjoint SO(3) coupling 
($\beta_A$) (see Section 2) will not alter the continuum 
limit of the mixed SU(2)-SO(3) action. From our work\cite{us1,us2,us3} 
on the mixed action at non-zero temperatures,
we, however, found the following surprising results:

\begin{enumerate} 
\item[{a]}] The transition remained second order in agreement with the 
Ising model exponents up to $\beta_A \approx 1.0$ but it became 
definitely first order for larger  $\beta_A$.  The order parameter for 
the deconfinement phase transition became nonzero discontinuously at 
this transition for larger $\beta_A$.

\item[{b]}] There was no evidence of a second separate bulk 
transition. 
\end{enumerate}

Similar results, indicative of apparent violation of universality, were also 
obtained with a variant Villain form for the SU(2) mixed action\cite{Ste} 
and for the $SO(3)$ lattice gauge theory\cite{DatGav}. 
Such qualitative violations {\it must be distinguished} from the quantitative 
violations of universality found earlier in\cite{BhaDas}.  
The latter disappear as the lattice spacing is reduced to zero 
and higher order corrections in 
$g_u^2$ are included\cite{MuSch,Gav,GavKarSat}. 
This will obviously not be the case for any qualitative 
disagreements like the one mentioned above for the universality of
the SU(2) deconfinement transition with the Ising model. 
While the change in the order
of the deconfinement phase transition was found on lattices
with a variety of different temporal sizes, only one transition was
found in each case at which the deconfinement order parameter
became non-zero.  This led to the suspicion that the bulk line
was misidentified and was actually a deconfinement line.  On the other
hand, the transition at larger $\a$ shifted very little, even as
the temporal lattice size was increased up to 16.  One expects the 
deconfinement transition to move to larger couplings as $\t$, i.e the 
temporal lattice size,  becomes larger.
A lack of shift was therefore more in tune with it being a bulk transition.
Direct numerical\cite{gavbulk} and analytical\cite{DatGav2}
evidences for the presence of the bulk transition were also
obtained.  A possible way out was to blame the small temporal extent of the
lattices used. Nevertheless, if a universal result exists in the 
$a \rightarrow 0$ limit, then  the tricritical point T, where the 
deconfinement phase transition changes its nature from second order 
(thin dotted lines in Fig. 1) to first order (thick dotted lines in 
Fig. 1), must not be seen in the continuum limit. Therefore, one 
expects that the tricritical point will move rapidly towards the upper 
right corner $(\f \rightarrow \infty, \a \rightarrow \infty)$ of Fig. 1 
as the lattice size increases.  In the continuum limit, the deconfinement 
transition will then be second order and its universality with Ising model 
will be recovered. However, in \cite{us3} we found 
that on going from $\t$ = 2 to 4 the tricritical point moved 
towards  lower values of the adjoint coupling. 
Studying the shapes of the histogram and invoking Polyakov loop 
effective potential arguments\cite{us3}, we  concluded 
that the tricritical point moved slowly upwards, if at all, on going from 
$N_{\tau} = 4 \rightarrow 6 \rightarrow 8$ lattices\footnote{Such
shifts of the tricritical points first towards lower values of $\beta_A$ 
on going from $N_{\tau}$= 2 to 4 and then towards higher values for 
$N_{\tau}$ =6, 8 were also observed in  \cite{Ste} where the Villain 
form (\ref{parfunv}) was simulated.}. These results are paradoxical 
and need explanation. In view of the above, i.e, the change in 
the nature of the deconfinement transition from second order to first order,
a qualitative violation in universality is a real possibility.  

In this paper, we take a different approach in an attempt to resolve these
paradoxes. While larger temporal lattices could perhaps be used
to solve them by brute force, a physical understanding of 
the universality of the continuum limit of the $SU(2)$ lattice gauge 
theory clearly requires an insight into the origin of 
the first order deconfinement transition and the tricritical point. 
In particular, one needs to 
a) identify the degrees of freedom which cause this unexpected change in 
the order of the deconfinement transition, and b) study whether these 
degrees of freedom are relevant or irrelevant in the continuum limit. 
Towards this end, we investigate the effect of the topological 
$Z_2$ degrees of freedom  associated with the $SO(3)$ part of the 
action (i.e, the term proportional to $\beta_A$ in (\ref{parfunv}))
on the deconfinement transition line in the
extended $(\f, \a)$ coupling plane. 
It is important to note that these $Z_2$ degrees of freedom defined 
below are different from those defined for SU(2) Wilson theory ($\beta_A = 0$) 
\cite{MacPet,Tom2,BroKes}. By adding  chemical potential
terms for the $Z_2$  monopoles and vortices, believed\cite{MacPet} to be 
irrelevant in the continuum limit, we show that no first order 
deconfinement phase transition exists if they both are suppressed. 
Thus the onset of the first order deconfinement seems to be due to the $Z_2$ 
topological degrees of freedom.  
In the past these $Z_2$ degrees of freedom have been studied in the 
context of bulk transitions in the extended model and cross over 
phenomenon along the Wilson axis $(\beta_A = 0)$ 
\cite{BroKes,CanHalSch,Tom1,Tom2}, the bulk transition along the $SO(3)$ 
gauge theory \cite{HalSch} and the $Z_2$ vortex transition along the 
$\beta_A = \infty$ axis. We propose to use them for the study of deconfinement
phase transition and for the entire $(\f, \a)$ plane.

We employ the Villain form of the mixed SU(2) action for our studies
below and find :

\begin{enumerate} 

\item The theory defined by mixed action is self-dual under the exchange 
of the fundamental and 
adjoint couplings in the absence of $Z_2$ magnetic monopoles.  

\item Further suppressing the $Z_2$ vortices completely reduces it to 
the standard Wilson action but 
with modified coupling; it thus can have only a second order deconfinement 
transition. 

\end{enumerate} 

\bigskip
\begin{center}
\bf 2. THE MODEL \\
\end{center}
\bigskip

The Villain form of the SU(2)-SO(3) mixed action is defined by 

\begin{eqnarray}
Z = \prod_{n,\mu} \sum_{\sigma_{p}(n) = \pm 1} \int dU_\mu(n)~~ 
\exp \left( {1 \over 2} \sum_p \left(\beta  + \beta_A \sigma_{p}\right) 
Tr_F U_p \right) ~~~.
\label{parfunv}
\end{eqnarray}

\noindent In (\ref{parfunv}) $U_{\mu}(n)$ are the SU(2) link variables, 
$\sigma_{p}(n)$ are the $Z_2$ plaquette fields. The first term in 
(\ref{parfunv}) describes the standard SU(2) Wilson action and the second 
term  describes the SO(3) action. 
 As mentioned earlier, this form was used in \cite{Ste} for 
studies similar to ours.  Indeed, apart from changes in numerical values 
of the couplings, the results were identical, leading to essentially the 
same phase diagram as in Fig. 1.

The main advantage of the Villain form is that 
the dynamical excitations corresponding to the $Z_2$ 
topological degrees of freedom, both magnetic and electric in nature,
become visible\cite{HalSch} in the partition function.  
In particular, it provides a better control over them and 
their contributions to any physical phenomenon. 

The $Z_2$ magnetic monopoles and $Z_2$ electric vortex charges are  obtained 
through the plaquette field $\sigma_p \equiv \sigma_{\mu\nu}$ as follows. 
We first define the $Z_2$ field strength tensor $F_{\mu\nu}$ by

\begin{eqnarray} 
\sigma_{\mu\nu}(n) \equiv \exp (i \pi F_{\mu\nu})
\label{z2fst}
\end{eqnarray} 

The $Z_2$ magnetic (electric) currents $\rho_c$ $(\rho_l)$ 
over a cube c (link l) located at a lattice site $n$ and oriented 
along $(\mu,\nu,\lambda)$ $(\mu)$  directions are defined as follows

\begin{eqnarray} 
\rho_c (n) \equiv \prod_{p \in \partial~ {c}} \sigma_p~ , ~~~~~~~~ 
\rho_l (n) \equiv \prod_{p \in {\tilde{\partial}}~ l} \sigma_p. 
\label{z2hch} 
\end{eqnarray}

\noindent In (\ref{z2hch}) $\partial$ and ${\tilde{\partial}}$ 
stand for boundaries and co-boundaries of the cube c and the link 
l respectively. 
We extract the standard electric and magnetic currents 
by the identification: 

\begin{eqnarray} 
\rho_l (n) \equiv \exp\left(i \pi J_{\mu}(n)\right),~~~~~ 
\rho_c (n) \equiv \exp\left(i \pi M_{\mu}(n)\right), 
\label{z2cur}
\end{eqnarray} 

\noindent In (\ref{z2cur}) the electric current $J_{\mu}(n)$ are 
on the link l and the magnetic current 
$M_{\mu}(n)$ is defined on the link $\mu$ which is orthogonal to the cube c. 
Using (\ref{z2fst}), (\ref{z2hch}) and (\ref{z2cur}), we recover 
the $Z_2$ Maxwell equations in the standard form: 

\begin{eqnarray}
\Delta_{\mu}F_{\mu\nu} (n) = J_{\nu} (n), ~~~  
\Delta_{\mu}{\tilde{F}}_{\mu\nu} (n) = M_{\nu} (n), ~~~  
\label{z2me} 
\end{eqnarray}

\noindent where, $\Delta_{\mu}(n)$ is the lattice difference 
operator in the $\mu^{th}$ direction. It is clear from (\ref{z2hch}) 
that the $Z_{2}$ magnetic and electric charge definitions are dual to 
each other. Note that the $Z_2$ monopoles defined by the second equation 
in (\ref{z2cur}) are different from those obtained by an abelian projection 
\cite{thooft} with respect to an SU(2) adjoint Higgs\footnote{e.g in the 
pure gauge theory  one could choose \cite{thooft} $\vec{\phi} \equiv 
\vec{F}_{12}$.}  field $\vec{\phi}$
with the abelian field strength tensor defined by: 

\begin{eqnarray}
F_{\mu\nu}^{U(1)} \equiv {\vec{\phi}.\vec{F}_{\mu\nu} \over |\vec{\phi}|} 
- {\vec{\phi} \over |\vec{\phi}|^{3}} \left(D_{\mu}\vec{\phi} \times D_\nu 
\vec{\phi}\right) .  
\label{abelian} 
\end{eqnarray}      

\noindent Here $\vec{F}_{\mu\nu}$ is the SU(2) field strength 
tensor derived from  $Tr U_{p}$. The U(1) topological 
magnetic current, derived from (\ref{abelian}),
and the corresponding U(1) magnetic monopoles are clearly different from 
the $Z_{2}$ monopoles described by (\ref{z2hch}).  It has been suggested 
\cite{thooft}  that the condensation of the former is perhaps responsible 
for confinement of color. On the other hand, the  $Z_2$ topological 
degrees of freedom defined in (\ref{z2hch}) are the lattice artifacts. 
We will show that they are responsible for changing 
the order of the deconfinement transition in the extended $(\beta,\beta_A)$
coupling plane.

In order to control the effects of the $Z_2$ topological degrees of 
freedom we add two terms proportional to their  $Z_2$ magnetic
and electric vortex charges in (\ref{parfunv}) :

\begin{eqnarray}  
Z_{\lambda,\gamma} = \prod_{n, \mu} \sum_{\sigma_{p}(n) = \pm 1} \int  
dU_\mu(n) \exp  \left( {1 \over 2} \sum_p \left(\beta  + \beta_A  
\sigma_{p}\right) Tr_F U_p + \lambda \sum_c  \rho_c + \gamma \sum_l\rho_l\right) ~.
\label{parfunl}
\end{eqnarray} 

\noindent Note that the formal continuum 
limit is expected to be left unchanged by the addition of $\lambda$ and 
$\gamma$ terms for all values of $\lambda$ and $\gamma$.
We discuss below the following cases:

\begin{enumerate}

\item $\gamma=0, \lambda = \infty$, i.e, a complete suppression of 
the $Z_2$ monopoles without affecting the $Z_2$ electric vortices. 

\item $\gamma = \infty, \lambda= 0$, i.e, a complete suppression 
of $Z_{2}$ electric vortices without affecting the $Z_2$ magnetic 
monopoles. 

\item $\gamma = \infty, \lambda = \infty$, i.e, complete suppression 
of all the $Z_2$ topological degrees of freedom. 

\end{enumerate}

\begin{center}
{\bf Case 1.} 
\end{center}
In the limit $\lambda = \infty$ at $\gamma =0$, 
one obtains the following constraint: 

\begin{eqnarray} 
\sigma_{c} = 1 ~~~~~~ \forall c
\label{con} 
\end{eqnarray} 

\noindent This amounts to complete suppression of point like 
magnetic monopoles. Note that these local constraints can still allow 
the global $Z_2$ magnetic fluxes arising due to the periodic  boundary 
conditions. While such boundary conditions in the spatial directions 
are not obligatory, they have to be enforced in the temporal directions. 
However, in what follows these global fluxes are irrelevant for our results.   
The constraints (\ref{con}) are the Bianchi identities in the absence of 
magnetic charges (see (\ref{z2cur}) and (\ref{z2me})). 
They  can be be trivially solved by introducing $Z_2$ link fields 
$\sigma_{\mu}(n)$, such that

\begin{eqnarray} 
\sigma_p \equiv \sigma_{\mu\nu}(n) = \sigma_{\mu}(n)\sigma_{\nu}(n+\mu)\sigma_{\mu}^{\dag}(n+\nu)\sigma_{\nu}^{\dag}(n)  
\label{sol} 
\end{eqnarray} 
\bigskip
\noindent The partition function in this limit therefore becomes

\begin{eqnarray}
Z_{\lambda \rightarrow \infty,\gamma} = \prod_{n,\mu} \sum_{\sigma_{\mu}(n) 
= \pm 1} \int  dU_\mu(n)~ \exp \Bigg( {1 \over 2} \sum_{p} \Big( \beta
 ~~~ + \hspace{3cm}~~~~~\nonumber \\
~~~~~~~~~~~\beta_A  \left( \sigma_{\mu}(n)\sigma_{\nu}(n+\mu)\sigma_{\mu}(n+\nu)\sigma_{\nu}(n)\right) \Big) Tr_F U_p \Bigg)~~  .
\label{inft1}
\end{eqnarray} 

\noindent It has a local $SU(2) \otimes Z_2$ gauge invariance: 

\begin{eqnarray} 
U_{\mu}(n) \rightarrow G(n) U_{\mu}(n) G^{-1}(n + \mu) ~~~~ G(n) \in SU(2) 
\nonumber \\ 
\sigma_{\mu}(n) \rightarrow \sigma_{\mu}(n) ~~~~~~~~~~~~~~~~~~~~~~~~
\label{su2z21}
\end{eqnarray} 
\begin{eqnarray}
\sigma_{\mu}(n) \rightarrow z(n) \sigma_{\mu}(n) z^{-1}(n + \mu) ~~~~ 
z(n) \in Z_2 \nonumber \\ 
U_{\mu}(n) \rightarrow U_{\mu}(n) ~~~~~~~~~~~~~~~~~~~~~~ 
\label{su2z22}   
\end{eqnarray} 

\noindent We now define a new SU(2) field on each link: 

\begin{eqnarray}
\tilde{U}_{\mu}(n) \equiv \sigma_{\mu}(n) U_{\mu}(n) 
\label{nvar} 
\end{eqnarray} 

Exploiting the SU(2) group invariance of the Haar  measure and the $Z_2$ 
nature of $\sigma$-variables the 
partition function (\ref{inft1}) in terms of new variables is:

\begin{eqnarray}
Z_{\lambda \rightarrow \infty,\gamma} = \prod_{n,\mu} \sum_{\sigma_{\mu}(n) 
= \pm 1} \int  
dU_\mu(n)~ \exp \Bigg( {1 \over 2} \sum_{p} \Big(\beta_A ~~~ + \hspace{3cm}~~~~~\nonumber \\ 
~~~~~~~~~~~\beta  
\left( \sigma_{\mu}(n)\sigma_{\nu}(n+\mu)\sigma_{\mu}(n+\nu)\sigma_{\nu}(n)\Big)
\right) Tr_F U_p \Bigg)~~.
\label{inft2}
\end{eqnarray} 

Eqs. (\ref{inft1}) and (\ref{inft2}) have exactly the same form 
with $\f \leftrightarrow \a $. Therefore, in the 
absence of $Z_2$ magnetic monopoles the physics of the extended model is 
self-dual under the exchange of the fundamental and adjoint couplings. 

Note that the self-duality arguments 
also go through for the SU(N) $(\forall N)$ extended Villain actions. 
The plaquette and the link fields  $\sigma_{\mu\nu}$ and $\sigma_{\mu}$ 
take values in the $Z_{N}$ group in that case.  $\rho_c$ is
clearly then complex and the action in (\ref{parfunl}) will
have to be modified by including its real part there.  However,
rest of the arguments are exactly the same as for the SU(2) case 
and again, in the limit of total suppression of $Z_N$ monopoles, the
physics will be self-dual under $\beta \leftrightarrow \beta_A$. 

\begin{center}
{\bf Case 2.} 
\end{center}
\bigskip 

The limit $\gamma = \infty, \lambda = 0$ corresponds to solving 
the constraint:  

\begin{eqnarray} 
\rho_l = 1 ~~~~~~~~~\forall~  l ~~.
\label{si}
\end{eqnarray} 

\noindent This limit suppresses 
all the $Z_2$ electric vortices. It is therefore dual to the $\lambda 
= \infty$ limit discussed above. Again, as discussed in Case 1 above 
the local constraints (\ref{si}) leave the global $Z_2$ electric 
fluxes due to the periodic boundary conditions unaffected. The solution 
of (\ref{si}) corresponds to  solving $\Delta_{\mu} F_{\mu\nu} =0$ 
in terms of the dual vector potentials and can be 
written as: 

\begin{eqnarray}
\sigma_{\mu\nu}(n) = {\rm exp} i \epsilon_{\mu\nu\rho\delta} \Delta_{\rho} 
\sigma_{\delta}(n) 
\label{dual}
\end{eqnarray} 
   
\noindent In (\ref{dual}) $\sigma_{\delta}(n)$ describes the dual 
vector potential on the link $\delta$ at lattice site n. 
Unlike in the case of the suppression of $Z_2$ magnetic 
charges suppression above, the suppression of $Z_2$ electric vortices 
does not lead to a self-dual model.

\bigskip 

\begin{center}
{\bf Case 3.}  
\end{center}
\bigskip 

The $\lambda = \gamma =  \infty$ limit 
corresponds to  complete suppression of both magnetic as well as 
electric $Z_2$ topological degrees of freedoms. The Maxwell eqns. 
in (\ref{z2me}) are now without electric or magnetic sources:

\begin{eqnarray}
\Delta_{\mu} F_{\mu\nu}(\sigma) = 0, ~~~~~~~ \Delta_{\mu} 
\tilde{F}_{\mu\nu}(\sigma) = 0. 
\label{2nd}
\end{eqnarray} 

\noindent The solution of the second of equation above, involving
the dual fields, is still given by eqn. (\ref{sol})). One thus has to solve 
for the remaining electric equation with the additional constraint of gauge
invariance of the $Z_2$ link fields $\sigma_\mu(n)$ which define the
$F_{\mu\nu}$.  Fixing the gauge symmetry (eqn. (\ref{su2z22})) such 
that $\Delta_{\mu} A_{\mu} = 0$, where the $Z_2$ vector potential
$A_\mu$ is related to $\sigma_\mu$ as $ \sigma_\mu = {\rm exp} (i A_\mu)$,
one can easily show that the gauge potential $A_{\mu}$ satisfies
the following equation:

\begin{eqnarray}
\Box A_{\mu}(n) = 0 ~~.~~
\label{3rd}
\end{eqnarray} 

\noindent It can be solved by a Fourier transform,
leading to a solution $\sigma_{\mu}(n)$ = constant, which 
can be $+1$ or $-1$ leading to $\sigma_{p} = 1$. Substituting back in 
(\ref{parfunl}), the extended action in this limit
reduces to the standard Wilson action with coupling $\beta + 
\beta_A$. It must therefore have a second order deconfinement transition
along the line $\beta + \beta_A = constant$, as known from the
results for $\beta_A = 0$.

\bigskip
\begin{center}
\bf 3. SUMMARY and DISCUSSION \\
\end{center}
\bigskip

Assuming the $Z_2$ topological degrees of freedom to be 
irrelevant in the continuum, as suggested by perturbative 
arguments, we have shown that the 
{\it qualitative} changes in the SU(2) deconfinement transition found 
earlier both by Monte Carlo as well as strong coupling expansions are 
likely to be due to the above degrees of freedom and hence lattice 
artifacts; suppressing them completely leads to the same physical result
for the deconfinement phase transition for all $\beta_A$, including large 
$\beta_A$. These qualitative changes in the SU(2) gauge theory 
on lattice are perhaps similar in spirit to the unphysical {\it confining} 
phase of U(1) compact lattice gauge theory.  The latter is also due to  
to the topological U(1) magnetic monopoles which are irrelevant in the 
continuum.  In order for the above universal results for the SU(2) 
deconfinement phase transition to be valid for the mixed action it 
is important to establish non-perturbatively that the $Z_2$ degrees 
of freedom are indeed irrelevant in the continuum limit. 
In the context of pure SU(2) lattice gauge theory it has been 
shown\cite{MacPet}  that the probability for the 
$Z_2$ monopoles excitation on a set of C cubes: 

\begin{eqnarray} 
< \prod_{c \in C} \theta(-\sigma_{c}) >_{Limit~  \beta 
\rightarrow \infty} ~~ \leq ~~ L_{1} \exp \left(- L_{2}~ \beta ~ C\right)  
\label{ineq}
\end{eqnarray} 

\noindent In (\ref{ineq}) $\theta(x) \equiv 1 (0)$  if x  $\ge$ 0 
(otherwise) and $L_{1}$, $L_{2}$ are constants.  
Depending on the value of $L_2$, the $Z_2$ monopoles could be extremely 
rare. Also, the $Z_2$ electric vortices are known to condense 
and give rise to a first order bulk transition shown by line $CG$ in Fig. 1. 
These arguments suggest that the original theory, i.e., $\gamma = \lambda 
=0$, should also exhibit universality for very large $N_{\tau}$.  Of
course, it is still necessary to extend the argument of Ref. \cite{MacPet}
to the entire $(\beta, \beta_A)$ plane for both the $Z_2$ monopoles
and vortices to be sure that this is indeed the case.  Moreover, the
curious coincidence of the bulk and deconfinement transition for the
case of $\gamma = \lambda=0$ for many different sizes of temporal
lattice sizes still remains unexplained, although one can now be more
hopeful that for $\it very$ large lattices at least the universality
will be respected, as it should be.

\bigskip 

\begin{center}
\bf 6.ACKNOWLEDGMENTS \\
\end{center}

\bigskip 

One of us (R.V.G.) wishes to thank Prof. Y. Iwasaki and Prof. A. Ukawa
for their invitation to visit the Center for Computational
Physics, University of Tsukuba, Japan, where most of this work was done.
The kind hospitality of the all the members of the center is gratefully
acknowledged.  M.M. would like to thank Prof. H. S. Sharatchandra for 
inviting him to The Institute of Mathematical Sciences, Chennai, 
India, where initial part of this work was done.


\newpage

\begin{thebibliography}{99}
\bibitem{thooft} G. t' Hooft Nucl. Phys.  {\bf B190} (1981) 455.
\bibitem{HalSch} G. Halliday, A. Schwimmer, Phys. Lett. {\bf 101B} (1981) 327 
and B. Lautrup, Phys. Rev. lett {\bf 47} (1981) 9, M. Creutz, Phys. Rev. Lett. 
{\bf 46} (1981) 1441. 
\bibitem{DasHel} R. Dashen, Urs M. Heller and H. Neuberger, Nucl. Phys.
{\bf B215[FS7]} (1983) 360.
\bibitem{AlbFly} J. M. Alberty, H. Flyvbjerg and B. Lautrup, Nucl. Phys. 
{\bf B220[FS8]} (1983) 61.
\bibitem{OgiHor} M. C. Ogilvie, A. Horowitz, Nucl. Phys. {\bf B215[FS7]} 
(1983) 249.
\bibitem{BhaCre} G. Bhanot and M. Creutz, Phys. Rev. {\bf D24} (1981) 3212.
\bibitem{gavbulk} R.V. Gavai, Nucl. Phys. {\bf B474 } (1996) 446
\bibitem{us1} R. V. Gavai, M. Grady and M. Mathur Nucl. Phys.
{\bf B423} (1994) 123.
\bibitem{us2} M. Mathur, R. V. Gavai, Nucl. Phys.  {\bf B448} (1995) 399;
Nucl. Phys. B (PS) 42 (1995) 490. 
\bibitem{us3} M. Mathur, R. V. Gavai, Phys. Rev. {\bf D56} (1997) 32.
\bibitem{FinHelKar} J. Fingberg, U. Heller and F. Karsch, Nucl. Phys. 
{\bf B392} (1993) 493.
\bibitem{EngFin} J. Engels, J. Fingberg and M. Weber, Nucl. Phys. 
{\bf B332} (1990) 737; \\
J. Engels, J. Fingberg and D. E. Miller, Nucl. Phys. 
{\bf B387} (1992) 501.
\bibitem{SveYaf} B. Svetitsky and L. G. Yaffe, Nucl. Phys. 
{\bf B210[FS6]} (1982) 423.
\bibitem{Ste} P. Stephenson, Talk given at EPS-HEP, Brussels, 
July August 95, hep-lat/9509070, heo-lat/9604008.
\bibitem{DatGav} Saumen Datta and R.V. Gavai,  Phys. Rev.
{\bf D57}(1998) 6618 .
\bibitem{EngSch} J. Engels and T. Scheideler, hep-lat/9610019. 
\bibitem{BhaDas} G. Bhanot and R. Dashen, Phys. Lett.
{\bf 113B} (1982) 299.
\bibitem{MuSch} K.-H. M\"utter and K. Schilling, Phys. Lett.
{\bf 121B} (1983) 267.
\bibitem{Gav} R.V. Gavai, Nucl. Phys. {\bf B215 [FS7]} (1983) 458.
\bibitem{GavKarSat} R.V. Gavai, F. Karsch and H. Satz, Nucl. Phys. {\bf B220 
[FS8]} (1983) 223.
\bibitem{DatGav2} Saumen Datta and R.V. Gavai,  Phys. Lett.
{\bf B392}(1997)172.
\bibitem{CanHalSch} L. Caneschi, I. G. Halliday and A. Schwimmer, Nucl. Phys.
{\bf B200 [FS4]} (1982) 409.
\bibitem{MacPet} G. Mack, V. B. Petkova, Ann. Phys. {\bf 125}, 117 (1980), 
Z. Phys. C {\bf 12}, 177 (1982).
\bibitem{Tom1} E. T. Tomboulis Phys. Lett.  {\bf 303B} (1993) 103.
\bibitem{Tom2} E. T. Tomboulis Phys. Lett.  {\bf 108B} (1982) 209.
\bibitem{BroKes} R. C. Brower, D. A. Kessler and H. Levine, Nucl. Phys.
{\bf B205[FS5]} (1982) 77.
\end{thebibliography}
\end{document}